\magnification=\magstep1
\input amstex
\documentstyle{amsppt}
\voffset-3pc
\baselineskip=24true pt

\topmatter
\title Nonstandard Feynman path integral for the
       harmonic oscillator \endtitle
\author Ken Loo \endauthor
\leftheadtext{Nonstandard Feynman Path integral for the
    harmonic oscillator}
\rightheadtext{Nonstandard Feynman Path integral for the
    harmonic oscillator}
\address PO Box 9160, Portland, OR.   97207
         \endaddress
\email look\@sdf.lonestar.org \endemail
\thanks The author would like to give thanks to Photine Tsoukalas
        \endthanks
\abstract Using Nonstandard Analysis, 
   we will provide a rigorous computation for the 
   harmonic oscillator Feynman 
   path integral.   
   The computation will be done 
   without having prior knowledge of the 
   classical path.  We will see that properties of classical
   physics falls out naturally from a purely quantum
   mechanical point of view.  We will assume that the reader
   is familiar with Nonstandard Analysis.   
\endabstract
\endtopmatter

\document
\define\n{{}^*\Bbb N - \Bbb N}

\define\fac#1#2{\left(\frac {m}{2\pi{}i\hbar\epsilon}\right)^
    {\frac{#1}2\left(#2 + 1 \right)}}

\define\summ#1#2#3#4{\frac {i\epsilon}{\hbar}\sum\limits_{j=1}
    ^{{#1}+1}\left[\frac {m}{2}\left(\frac {{#2} - {#3}}
    {\epsilon}\right)^2\! #4\right]}

\define\pd#1#2{\dfrac{\partial#1}{\partial#2}}
\define\spd#1#2{\tfrac{\partial#1}{\partial#2}}
\define\ham1#1#2{\dfrac{-\hbar^2}{2m}\Delta_{#1} #2}

\subhead
{\bf I. Introduction}\endsubhead

In quantum mechanics, we are interested in finding the wave function
which satisfies Schrodinger's equation.  Equivalently, we can find
the propagator or integral kernel $K(q,q_0,t)$ which satisfies
$$\align
  {}&i\hbar\pd {K(q,q_0,t)}{t} = \left[\ham1 {q}{+V(q)}\right]K(q,q_0,t),
    \tag1.1 \\
  {}&K(q,q_0,0) = \delta{}(q - q_o),\quad q,q_0\in\Bbb R{}^d.
\endalign
$$
Formally, the wave function is related to the propagator via
$$\align
  {}&\varphi{}(q,t) = \int_{-\infty}^{+\infty}
  K(q,q_0,t)\phi{}(q_0)\,dq_0\,{} = 
  \Bigg<K(q,q_0,t),\,\phi{}(q_0)\Bigg>, 
\endalign
$$
with boundary condition $\varphi(q,0) = \phi{}(q)$.  
In Feynman's formulation of quantum mechanics,
he proposed that the propagator is given
by a functional integral, also referred to as Feynman path integral or
just path integral
in physics literature,
$$\align
  {}&K(q,q_0,t)= \int_{x(0) = q_0}^{x(t) = q}
     \text{exp}\left\{\dfrac{iS\left[x\left(s\right)\right]}{\hbar}\right\}
     dx(s) = \tag1.3\\
  {}& \lim_{n\to\infty}\int\limits_{\Bbb R^{dn}}
   w_{d,n}\text{ exp }\left[\dfrac{i\epsilon}{\hbar}
  S\left\{\vec x_{n+1}\dots\vec x_{0}\right\}\right]
  \,d\vec x_1\dots d\vec x_n, \quad\text{where}\\
  {}&x_0=q_0, x_{n+1}=q,
     \epsilon =\dfrac{t}{n}, \\
   {}&w_{d,n} = \left(\dfrac{m}{2i\pi\hbar\epsilon}\right)^
   \frac{d(n+1)}{2}, \\
   {}&S\left\{\vec x_{n+1}\dots\vec x_{0}\right\} =
   \sum_{j=1}^{n+1}\left[\dfrac{m}{2}
   \left(\dfrac{\vec x_{j} - \vec x_{j-1}}{\epsilon}\right)^2 -
   V\left(\vec x_j\right)
   \right].
\endalign
$$
The integrals in the second line of (1.3) are $d$-dimensional improper 
Riemann integrals.
The integral in the first equality 
is the purely formal path integral which integrates
over all paths
$x\left(s\right), 0\leq{}s{}\leq{}t$, with $x\left(0\right) = q_0$, and
$x\left(t\right) = q$.  The quantity $S\left[x\left(s\right)\right]$ is
the action integral
$$ S\left[x\left(s\right)\right] = \int_0^t L\left(x,\dot x\right)\, ds,
\quad  L\left(x,\dot x\right) = \dfrac{m}{2}\dot x - V(x)\, . \tag1.4
$$
The motivation for the notation of the formal path integral
is that as $n$ goes to infinity,
the sum in the exponent becomes the action integral.
In this manner, Feynman was able to deduce Classical Mechanics 
from Quantum 
Mechanics through the action integral and $\hbar\to 0$.
In this paper, we will compute the harmonic oscillator
path integral without prior knowledge of classical physics.  
It turns out that properties of classical physics naturally
falls out of the computation without taking $\hbar\to 0$.  
The computation shows that the path integral separates into
the product of two quantities, one independent of $\hbar$,
the other dependent on $\hbar$.  The quantity which is 
independent of $\hbar$ contains properties of classical physics.
Thus, in some sense
we are deviating from the standard interpretation
of the path integral being a bridge between
Quantum Mechanics and Classical Mechanics via the
action integral and $\hbar$.  We are considering
Quantum Mechanics as purely Quantum Mechanics and 
extracting properties of Classical Mechanics without
prior knowledge of Classical Physics.

Mathematically, the formal integration over paths 
can not be a rigorously well defined 
measure theoretic integration because of the oscillatory
nature of the integrand(see [1] and [2]).  
A popular technique to make sense of the (1.3)
is to replace $t$ by $-it$
and use the Wiener integral(see [3] and [4]).  

In Nonstandard Analysis, we have that
$\lim_{n\to\infty} a_n = a$ iff ${}^*a_{\omega} \approx a$
for any infinite natural number $\omega\in\n$, with
$\left\{{}^*a_m\right\}_{m\in{}^*\Bbb N}$
being the $*$-extension of
$\left\{a_n\right\}_{n\in\Bbb N}$, and $\approx$ means that $a_\omega +
h_\omega = a$ where $h_\omega$ an infinitesimal.
We can use Nonstandard Analysis to define the path integral; 
the standard part can replace the limit in (1.3).  
Using Nonstandard Analysis to replace the limit in (1.3) 
is not a new concept(see [5], [6] and references within), doing so 
partially solves the problems of the Feynman path integral
on the propagator.  

We can redefine (1.3) in the following
manner: 
let $\omega\in{}^*\Bbb N, m,t\in\Bbb R^+,
  \epsilon =
  \frac{t}{\omega},
     {}^*V(x):{}^*\Bbb R^d \rightarrow {}^*\Bbb R$
      be an internal function, and
      $x_0 = q_0, x_{\omega +1}
     =q$ be fixed points in $\Bbb R^d$.  We call the expression:
  $$
       \int\limits_{{}^*\Bbb R^{d\omega}}
        \fac{d}{\omega}\text{exp}\left[\summ{\omega}{x_j}{x_{j-1}}
        {-{}^*V(x_j)}\right]
      dx_1\dots dx_{\omega}\,  \tag1.5
      $$
   an internal 
   functional integral.  In (1.5), all integrals are 
   $*$-transformed improper Riemann integrals.  
   In particular, if 
   for all $\omega\in\n$, the standard part of the 
   internal functional integral
   exists and it is independent of the choice of $\omega$, we call the
   standard part a standard functional integral or Feynman
   path integral and denote it by
  $$\align
      {}&st\!\!\!\int\limits_{{}^*\Bbb R^{d\omega}}
        \fac{d}{\omega}\!\!\!\text{exp}\left[\summ{\omega}{x_j}{x_{j-1}}
        {-{}^*V(x_j)}\right]
      dx_1\dots dx_{\omega} .\tag1.6 \\
   \endalign
   $$

Equation (1.6) is just
a Nonstandard Analysis way of saying that the limit in (1.3) 
exists.  There still remain the problem of for which class
of potentials $V$ the expression (1.5) exists and whether
(1.6) actually produces the propagator.  

We will demonstrate the usage of (1.5) and (1.6)
on the $d$-dimensional harmonic oscillator path integral.
The harmonic oscillator plays a major role in quantum field theory
and the recent advances due to Duru and Kleinert
in the coulomb potential path integral(see [7] and references within).
The harmonic
oscillator carries the potential
$V\left(x\right) = \dfrac{m\lambda^2}{2}x^2$,
its internal functional integral is:
 $$
      \int\limits_{{}^*\Bbb R^{d\omega}}
        \fac{d}{\omega}\text{exp}\left[\summ{\omega}{x_j}{x_{j-1}}
       {-\dfrac{m}2\lambda^2 x_j^2}\right]
      dx_1\dots dx_{\omega}\, . \tag1.7
 $$
Notice that each $d$-dimensional integral factors into
$d$ products of one dimensional integrals, thous we shall
compute the $1$-dimensional harmonic oscillator internal
functional
integral.

\subhead
{\bf II. The Harmonic Oscillator}\endsubhead
There are many ways to compute the harmonic oscillator
path integral(see [7], [8], [9] and references within), 
we will use a method of 
computation similar to that of [8].
We differ from the popular techniques in that we will do the
computation without prior knowledge of the classical
path of the harmonic oscillator and we will rigorously 
do the computation
with Nonstandard Analysis.  
As pointed out earlier, it turns out that
properties of Classical Physics falls out naturally from a purely
Quantum Mechanical derivation.  

It is well known that for $0 < t < \dfrac{\pi}{\lambda}$,
the propagator for the $1$-dimensional harmonic
oscillator is
$$K\left(q,q_0,t\right) = \left(\dfrac{m}{2\pi{}i\hbar}\right)^{\frac{1}2}
\sqrt{\dfrac{\lambda}
{\sin\lambda{}t}}
\text{exp}\left\{\dfrac{im}{\hbar}\dfrac{\lambda}{\sin\lambda{}t}
\left[\left({q_0}^2 + q^2\right)\cos\lambda{}t - 2qq_0\right]
\right\}.\tag2.1 $$
Equation (2.1) carries 
a singularity at $t = \dfrac{\pi}{\lambda}$.  Singularity
like that of (2.1) can in fact be given rigorous  
mathematical meaning if $K\left(q,q_0,t\right)$
is interpreted as a distribution(see [10]).  
Due to the form of (2.1), we would not expect the standard
functional integral to exist at $t = \dfrac{\pi}{\lambda}$ as a
function.  

The popular method to compute the $1$-dimensional harmonic oscillator
Feynman path integral is by writing $S\left[x\left(s\right)\right] =
S\left[x^{cl}\left(s\right) + \delta{}x\left(s\right)\right]$, where
$x^{cl}\left(s\right)$ is the classical path of the harmonic oscillator
which satisfies the equation of motions
$\ddot x^{cl}\left(s\right) = -\lambda^2{}x^{cl}\left(s\right)$, with the
boundary condition $x^{cl}\left(0\right) = q_0, x^{cl}\left(t\right) = q$.
Namely, $x^{cl}\left(s\right) = \dfrac{q\sin\lambda{}s +
q_0\sin\lambda\left(t-s\right)}{\sin\lambda{}t}$.  The action integral becomes
$$\align
  {}&\int_0^t \dfrac{m}2 \left[\left(\dot x^{cl}\right)^2 -
    \lambda^2\left(x^{cl}\right)^2\right]\,ds +
    \int_0^t \dfrac{m}2 \left[\left(\delta\dot x\right)^2 -
    \lambda^2\left(\delta x\right)^2\right]\,ds + \tag2.2\\
  {}&\int_0^t m \left(\dot x^{cl}\delta\dot x\ -
     \lambda^2 x^{cl}\delta x\right)\,ds\, .
\endalign
$$
Using $\delta x\left(0\right) = 0 = \delta x\left(t\right)$
and the equation of motions of $x^{cl}$, the last integral
is $0$ after an integration by parts.
Integrating by parts on the first integral and using the equation
of motions of $x^{cl}$, the first integral becomes
$\dfrac{m}2{}x^{cl}\dot x^{cl}
\bigg|_0^t = \dfrac{m\lambda}{2\sin\lambda{}t}\left[\left(q^2 + q_0^2\right)
\cos{}t - 2qq_0\right]$.  Without much concern on the existence and meaning
of the path integral, we can write
$$\align
     {}&\int_{x(0) = q_0}^{x(t) = q}
     \text{exp}\left\{\dfrac{iS\left[x\left(s\right)\right]}{\hbar}\right\}
     dx(s) =  \tag2.3 \\
     {}&\text{exp}\left\{\dfrac{im\lambda}{2\hbar\sin\lambda{}t}
     \left[\left(q^2 + q_0^2\right)
     \cos{}t - 2qq_0\right]\right\}
     \int_{\delta{}x(0) = 0}^{\delta{}x(t) = 0}
     \text{exp}\left\{\dfrac{iS\left[\delta{}x\left(s\right)\right]}
     {\hbar}\right\}
     d\delta{}x(s)\, .
   \endalign
$$
The path integral on the right-hand side is the quantum fluctuation;
the integral is over all paths which starts from $0$ at $s=0$, and end at $0$
at $s=t$.  We leave it to the reader to look up
the computation of the quantum fluctuation in the literature.

We will give a rigorous
treatment of (2.3) by using the time-sliced
internal path integral in (1.5) and (1.6).
In our work, we do not start with having knowledge of the
classical path.
We will start with an arbitrary bounded path $w(s)$ which satisfies
$w\left(0\right) = q_0, w\left(t\right) = q$, and separate the
propagator into a product of  classical and quantum amplitudes.
In this approach,
we will see that the classical contribution actually comes from
Quantum Mechanics without prior knowledge of Classical Mechanics.

To shorten the notation, we write
$$\align
   {}&\summ{n}{x_j}{x_{j-1}}{-\dfrac{m}2\lambda^2 x_j^2} = \tag2.4\\
   {}&\left(\dfrac {im}{2\hbar\epsilon}\right)
   \left[x_0^2 - 2x_0x_1 + x_{n+1}^2 - 2x_nx_{n+1} +
   \sum\limits_{j = 1}^{n} 2x_j^2 - 
   \sum\limits_{j = 1}^{n} 2x_jx_{j-1} -
   \epsilon^2\lambda^2 \sum\limits_{j = 1}^{n+1}x_j^2
   \right] = \\
\endalign
$$
$$\align
   {}&\left(\dfrac {im}{2\hbar\epsilon}\right)
   x^t\Bigg\{
   \left(\!\!\!\!\matrix\format\quad\r&\quad\r&\quad\r&\quad\r&\quad\r\\
   1&-1&0&\hdots&0\\
   -1&{ }&{ }&{ }&\vdots\\
   0&{ }& 0 &{ }&0\\
   \vdots&{ }&{ }&{ }&-1\\
   0&\hdots&{ }&-1&1\\
    \endmatrix\right)  \, -  \, \\ 
   {}&\epsilon^2\lambda^2
   \left(\matrix
    0&\hdots&{ }&{ }&{ }&\hdots&\hdots&0\\
    0&1&0&\hdots&{ }&{ }&\hdots&0\\
    \vdots&\ddots&\ddots&\ddots&\ddots&{ }&{ }&\vdots\\
    { }&{ }&\ddots&\ddots&\ddots&{ }&{ }&{ }\\
    { }&{ }&{ }&\ddots&\ddots&\ddots&{ }&{ }\\
    \vdots&{ }&{ }&{ }&\ddots&\ddots&{ }&\vdots\\
    0&\hdots&{ }&{ }&\hdots&0&1&0\\
    0&\hdots&{ }&{ }&{ }&\hdots&0&1\\
    \endmatrix\right)\, + \\
   {}&\left(\matrix
   0&\hdots&\hdots&\hdots&\hdots&\hdots&\hdots&\hdots&\hdots&0\\
   0&2&-1&0&\hdots&\hdots&\hdots&\hdots&\hdots&0 \\
   \vdots&-1 & 2 &-1&0&\hdots&\hdots & \hdots&\hdots &0 \\
    \vdots&0&-1&2&-1& 0 &\hdots&\hdots&\hdots&0 \\
    \vdots&\vdots&\ddots&\ddots&\ddots&\ddots&\ddots&\ddots&{ }&\vdots \\
    \vdots&\vdots&{ }&\ddots&\ddots&\ddots&\ddots&\ddots&\ddots&\vdots\\
    0&{ }&{ }&\ddots&\ddots&\ddots&\ddots&\ddots&{ }&{ }\\
    0 & \hdots&\hdots & \hdots &\hdots&0&-1 & 2 & -1 &0\\
    0 & \hdots& \hdots& \hdots & \hdots&\hdots&0&-1&2 &0\\
    0&\hdots&\hdots&\hdots&\hdots&\hdots&\hdots&\hdots&\hdots&0\\
    \endmatrix\right)\Bigg\}x = 
   \left(\dfrac {im}{2\hbar\epsilon}\right)
   \left(x^tT_nx\right), 
\endalign
$$
where $T_n$ is the $(n+2)$ by $(n+2)$ symmetric
matrix, 
$$T_n = \left(\!\!\!\!\matrix\format\quad\r&\quad\r&\quad\r&\quad\r&\quad\r\\
   1&-1&0&\hdots&0\\
  -1&{ }&{ }&{ }&\vdots\\
   0&{ }& S_n &{ }&0\\
   \vdots&{ }&{ }&{ }&-1\\
   0&\hdots&{ }&-1&1 - \epsilon^2\lambda^2\\
    \endmatrix\right),\tag2.5
$$
with $S_n$
being the $n$ by $n$ symmetric matrix
$
   S_n = A_n - \epsilon^2\lambda^2{}B_n 
$, where 
$$\align
   {}&A_n = \left(\matrix
   2&-1&0&\hdots&\hdots&\hdots&\hdots&0 \\
   -1 & 2 &-1&0&\hdots&\hdots & \hdots &0 \\
    0&-1&2&-1&\hdots&\hdots&\hdots&0 \\
    \vdots&\ddots&\ddots&\ddots&\ddots&\ddots&{ }&\vdots \\
    \vdots&{ }&\ddots&\ddots&\ddots&\ddots&\ddots&\vdots\\
    0&{ }&{ }&\ddots&\ddots&\ddots&\ddots&0\\
    0 & \hdots & \hdots &\hdots&0&-1 & 2 & -1 \\
    0 & \hdots & \hdots & \hdots&\hdots&0&-1&2
    \endmatrix\right), \tag2.6\\
\endalign
$$
$$\align
   {}&B_n = \left(\matrix
    1&0&\hdots&{ }&{ }&{ }&\hdots&0\\
    0&1&0&\hdots&{ }&{ }&\hdots&0\\
    \vdots&\ddots&\ddots&\ddots&{ }&{ }&{ }&\vdots\\
    { }&{ }&\ddots&\ddots&\ddots&{ }&{ }&{ }\\
    { }&{ }&{ }&\ddots&\ddots&\ddots&{ }&{ }\\
    \vdots&{ }&{ }&{ }&\ddots&\ddots&{ }&\vdots\\
    0&\hdots&{ }&{ }&\hdots&0&1&0\\
    0&\hdots&{ }&{ }&{ }&\hdots&0&1\\
\endmatrix\right)\,
\endalign
$$
and 
$x$ is the column vector
$$
   x = \left(\matrix
   x_0\\
   x_1\\
   \vdots\\
   x_n\\
   x_{n+1}\\
  \endmatrix\right)\, . \tag2.7
$$
For notation convenience, we will use a bar instead of $*$ to indicate
the $*$-transforms of matrices, determinate of matrices, and vectors.

We are interested in knowing when the internal functional integral of
the harmonic oscillator
exists.  From (2.1), it would be
reasonable to postulate the existence of the functional integral for
$t < \dfrac{\pi}{\lambda}$. Indeed, this turns out to be the case.
For $t < \dfrac{\pi}{\lambda}, \bar S_\omega$(the $*$-transform of $S_n$
with $n=\omega\in\n$) turns out to be $*$-positive definite,
which allows us to actually compute the integrals.

\proclaim{\bf Proposition 2.1} For $0 < t < \sqrt{
      \dfrac{n^2\pi^2}{\lambda^2\left(n+1\right)^2}
      \left[1-\dfrac{\pi^2}{12\left(n+1\right)^2}\right]
      },\,
S_n$ is positive definite.
\endproclaim

\demo{Proof} An elementary computation shows that for $k = 1,2,\dots ,n$,
$$  A_n\left(\matrix
    \sin\frac{k\pi}{n+1}\\
    \sin\frac{2k\pi}{n+1}\\
    \vdots \\
    \vdots \\
    \vdots \\
    \vdots \\
    \sin\frac{nk\pi}{n+1}\\
    \endmatrix\right) = \left(2-2\cos\frac{k\pi}{n+1}\right)\left(\matrix
    \sin\frac{k\pi}{n+1}\\
    \sin\frac{2k\pi}{n+1}\\
    \vdots \\
    \vdots \\
    \vdots \\
    \vdots \\
    \sin\frac{nk\pi}{n+1}\\
    \endmatrix\right)\, . \tag2.8 
$$
Hence, the $n$ distinct eigenvalues of $S_n\text{ are }2-
2\cos\left(\frac{k\pi}{n+1}\right) -
\lambda^2\left(\frac{t}{n}\right)^2$. To show that $S_n$ is positive definite,
it is enough to find the values of $t$ for which the eigenvalues are positive,
or $\cos\left(\frac{k\pi}{n+1}
\right) < 1 - \dfrac{\lambda^2t^2}{2n^2}$.  Since
$\cos\left(\frac{k\pi}{n+1}\right)\leq\cos\left(\frac{\pi}{n+1}\right)
\text{ for } k = 1, 2,\dots ,n$, it is enough to find $t$ for which
$\cos\left(\frac{\pi}{n+1}
\right) < 1 - \dfrac{\lambda^2t^2}{2n^2}$.  By Taylor expanding
$\cos\left(\frac{\pi}{n+1}\right)$ about $0$ for the first 3 nonzero
terms, we have
$$\align
  {}&\cos\left(\frac{\pi}{n+1}\right) < 1 - \dfrac{\lambda^2t^2}{2n^2}\quad
  \Leftrightarrow\quad
  1-\dfrac{\pi^2}{2\left(n+1\right)^2} +
  \dfrac{\pi^4\cos\eta}{4!\left(n+1\right)^4} < 1 - \dfrac{\lambda^2t^2}{2n^2}
  \quad \Leftrightarrow \tag2.9\\
  {}&\quad t^2 < \dfrac{\pi^2{}n^2}{\lambda^2\left(n+1\right)^2}
  \left[1 - \dfrac{\pi^2\cos\eta}{12\left(n+1\right)^2}\right],\quad
\endalign
$$
where $0 < \eta \leq \dfrac{\pi}{n+1}$.
When $t < \sqrt{
      \dfrac{n^2\pi^2}{\lambda^2\left(n+1\right)^2}
      \left[1-\dfrac{\pi^2}{12\left(n+1\right)^2}\right]
      },$ we have 
      $t^2 < \dfrac{\pi^2{}n^2}{\lambda^2\left(n+1\right)^2}
  \left[1 - \dfrac{\pi^2\cos\eta}{12\left(n+1\right)^2}\right]$. \qed
\enddemo

\proclaim{\bf Theorem 2.2} Let $t\in\Bbb R$ and
$0 < t < \dfrac{\pi}{\lambda}$.  For any  
$\omega\in{}^*\Bbb N - \Bbb N, \bar S_{\omega}$ is
positive definite in the $*$-transformed sense.
\endproclaim

\demo{Proof} $*$-transforming Proposition 2.1 and setting $n = \omega$,
we have that $\bar S_\omega$ is positive definite when
$$
      0 < t < \sqrt{
      \dfrac{\omega^2\pi^2}{\lambda^2\left(\omega{}+1\right)^2}
      \left[1-\dfrac{\pi}{12\left(\omega{}+1\right)^2}\right]
      } = \dfrac{\pi}{\lambda} + h\, , \tag2.10
$$
where $h$ is infinitesimal.  
When $t$ is standard and $0 < t < \dfrac{\pi}{\lambda}$,
(2.10) holds.  
\qed
\enddemo

From here on, let us take $0 < t <\dfrac{\pi}{\lambda}$.  
We will now proceed to separate the functional integral into a classical
part and a quantum fluctuation part.
Suppose $w\left(s\right)$ is an arbitrary path
with $|w\left(s\right)|
< \infty$ for $0 \leq s \leq t$.  Furthermore, 
let $w\left(0\right)=q_0$, and $w\left(t\right)=q$.
We make the substitution $x_j =
w\left(\frac {jt}{n+1}\right) + y_j = w_j + y_j$
(notice that $y_0 = 0 = y_{n+1}$ since $w\left(0\right) = x_0 = q_0$
and $w\left(t\right) = x_{n+1} = q$).  Using the fact that $T_n$ is
symmetric, we have 
$$\align
   {}&x^tT_nx = \left(y + w\right)^tT_n
     \left(y + w\right) =
     w^tT_nw + y^tT_ny + w^tT_ny + y^tT_nw = \tag2.11 \\
   {}&w^tT_nw + y^tT_ny +
     \left(T_nw\right)^{t}y + \left(w^{t}T_ny\right)^{t} =
     w^tT_nw + y^tT_ny +
     \left(T_nw\right)^{t}y + w^{t}T_ny = \\
   {}&w^tT_nw + y^tT_ny +
     2\left(T_nw\right)^ty
\endalign
$$
By using $y_0 = 0 = y_{n+1}$ and writing $T_n$ as
$$ T_n = 
   \left(\!\!\!\!\matrix\format\quad\r&\quad\r&\quad\r&\quad\r&\quad\r\\
   0&0&0&\hdots&0\\
   0&{ }&{ }&{ }&\vdots\\
   0&{ }& S_n &{ }&0\\
   \vdots&{ }&{ }&{ }&0\\
   0&\hdots&{ }&0&0\\
    \endmatrix\right) +
   \left(\!\!\!\!\matrix\format\quad\r&\quad\r&\quad\r&\quad\r&\quad\r\\
   1&-1&0&\hdots&0\\
  -1&{ }&{ }&{ }&\vdots\\
   0&{ }& 0 &{ }&0\\
   \vdots&{ }&{ }&{ }&-1\\
   0&\hdots&{ }&-1&1 - \epsilon^2\lambda^2\\
    \endmatrix\right),\tag2.12
$$
we obtain
$$x^tT_nx = w^tT_nw + y^tT_ny +
     2\left(T_nw\right)^ty = 
   w^tT_nw + \hat y^tS_n\hat y +
     2\rho^t\hat y\, ,\tag2.13
$$ where 
$$\align
  {}&y = \left(\matrix
     0\\
    y_1\\
    \vdots\\
    y_n\\
      0\\
   \endmatrix\right), \quad
   \hat y = \left(\matrix
    y_1\\
    \vdots\\
     y_n
   \endmatrix\right), \quad
   w = \left(\matrix
     w_0\\
     w_1\\
      \vdots\\
     w_{n}\\
     w_{n+1}\\
   \endmatrix\right), \quad
 \text{and}\tag2.14\\
 {}&\rho = S_n\left(\matrix
       w_1\\
       w_2\\
       \vdots\\
       w_{n-1}\\
       w_n\\
   \endmatrix\right) - \left(\matrix
       w_0\\
       0\\
       \vdots\\
       0\\
       w_{n+1} \\
   \endmatrix\right)\, = S_n\hat w - \hat{\hat w} .
\endalign
$$
We then have the following:
\proclaim{Lemma 2.3} Under the assumption on $t$,
let 
$\epsilon = \dfrac{t}{n}$, then
$$\align
      {}&\int\limits_{\Bbb R^n}
        \fac{1}{n}\text{exp}\left[\summ{n}{x_j}{x_{j-1}}
       {-\dfrac{m}2\lambda^2 x_j^2}\right]
      dx_1\dots dx_{n} = \tag2.15 \\
        {}&\text{exp}\left[\dfrac{im}{\hbar\epsilon}
        \left(w^tT_nw -
         \rho^tS_n^{-1}\rho\right)\right]
       \left(\dfrac{m}{2\pi{}i\hbar\epsilon}\right)^{\frac1{2}}
         \sqrt{\dfrac1{detS_n}}\, .
    \endalign
 $$
\endproclaim

\demo{Proof}
Since $S_n$ is positive definite, it is invertible.
Since $S_n$ is symmetric, the following holds
$$
  \hat y^tS_n\hat y+2\rho^t\hat y
   = \left(\hat y + S_n^{-1}\rho\right)^tS_n
  \left(\hat y + S_n^{-1}\rho\right) -
   \rho^tS^{-1}\rho. \tag2.16$$ 
Using our shorten notation in $(2.13)$ and (2.16), 
the integrals in (2.15) is equivalent to 
$$
        \text{exp}\left[\dfrac{im}{\hbar\epsilon}
        \left(w^tT_nw -
         \rho^tS_n^{-1}\rho\right)\right]
     \int\limits_{\Bbb R^n} \fac{1}{n}\!\!\!\!\!\!
     \text{exp}\left(\dfrac{im}{2\hbar\epsilon}z^tS_nz\right)\,
       dz_1\dots dz_n\, .\tag2.17
$$
In obtaining (2.17), we performed the change of
variables $x_j = w_j + y_j$, and then
from $y_j + \left(S_n^{-1}\rho\right)_j = z_j$.  
We get (2.15) after diagonalizing $S_n$ 
and doing the decoupled integrals.\qed
\enddemo

\proclaim{\bf Corollary 2.4} 
Under the previous definition of 
$w\left(s\right)$, 
The $1$-dimensional harmonic
oscillator internal functional integral is
well defined and it is equal to
$$       \text{exp}\left[\dfrac{im}{\hbar\epsilon}
        \left(\bar w^t\bar T_\omega\bar w -
         \bar\rho^t\bar S_\omega^{-1}\bar\rho\right)\right]
       \left(\dfrac{m}{2\pi{}i\hbar\epsilon}\right)^{\frac1{2}}
         \sqrt{\dfrac1{det\bar S_\omega}}\, .\tag2.18
$$
where the bars denote $*$-transform.
\endproclaim

\demo{Proof} This is just the 
$*$-transform lemma 2.3.\qed
\enddemo

\remark{\bf Remark 2.1} 
There is no restriction on the choice of the path $w$ except that
it starts at $q_0$ and ends at $q$.  We will show that the exponential part
of (2.18) 
turns out to be the classical amplitude of the previous formal calculation
of the propagator and the other factor is the quantum fluctuation.  If we
choose $\bar w$ to be the $*$-transform of the classical path, it can be shown
that each entry in $\bar\rho$ is infinitesimal, and
$\dfrac {\bar w^t\bar T_{\omega}\bar w}{\epsilon}$ is infinitesimally close to
$
\dfrac{\lambda}{\sin\lambda{}t}
\left[\left({q_0}^2 + q^2\right)\cos\lambda{}t - 2qq_0\right]
$.
\endremark

There are many techniques to compute the quantum fluctuation
$\dsize\lim_{n\to\infty}
\sqrt{\dfrac1{\epsilon detS_n}}$ in the literature; we present a
rigorous method to compute the limit with Nonstandard Analysis by computing
$st\sqrt{\dfrac1{\epsilon det\bar S_{\omega}}}$, for $\omega\in\n$.

\proclaim{\bf Proposition 2.5} Let 
$\epsilon = \dfrac
{t}{n}$.  Denote $A_{j,n}$ and $C_{j,n}$ with $0 < j \leq n$ 
to be the 
$j$ by $j$ matrices given by 
$$\align
   {}&A_{j,n} = \left(\matrix
   2&-1&0&\hdots&\hdots&\hdots&\hdots&0 \\
   -1 & 2 &-1&0&\hdots&\hdots & \hdots &0 \\
    0&-1&2&-1&\hdots&\hdots&\hdots&0 \\
    \vdots&\ddots&\ddots&\ddots&\ddots&\ddots&{ }&\vdots \\
    \vdots&{ }&\ddots&\ddots&\ddots&\ddots&\ddots&\vdots\\
    0&{ }&{ }&\ddots&\ddots&\ddots&\ddots&0\\
    0 & \hdots & \hdots &\hdots&0&-1 & 2 & -1 \\
    0 & \hdots & \hdots & \hdots&\hdots&0&-1&1
    \endmatrix\right), \text{ and } \tag2.19\\
    {}&C_{j,n} = \left(\matrix
    1&0&\hdots&{ }&{ }&{ }&\hdots&0\\
    0&1&0&\hdots&{ }&{ }&\hdots&0\\
    \vdots&\ddots&\ddots&\ddots&{ }&{ }&{ }&\vdots\\
    { }&{ }&\ddots&\ddots&\ddots&{ }&{ }&{ }\\
    { }&{ }&{ }&\ddots&\ddots&\ddots&{ }&{ }\\
    \vdots&{ }&{ }&{ }&\ddots&\ddots&{ }&\vdots\\
    0&\hdots&{ }&{ }&\hdots&0&1&0\\
    0&\hdots&{ }&{ }&{ }&\hdots&0&0\\
\endmatrix\right)\, .
\endalign
$$
Define $D_{j,n} = det\left\|A_{j,n} -
    \epsilon^2\lambda^2{}C_n\right\|.  
$
After $*$-transforming, we have that for 
$\omega\in\n$,
$\bar D_{k,\omega}\approx
{}^*\!\!\cos(\dfrac{kt\lambda}{\omega})={}^*\!\!\cos(k\epsilon\lambda)$.
\endproclaim

\demo{Proof}
For $k = 1$, $A_{1,\omega} = (1)$, $C_{1,\omega} = (0)$ and
$D_{1,\omega} = 1$.
For $k = 2$, 
$$
  \bar A_{2,\omega} =  
   \left(\matrix
   2&-1 \\
   -1&1
\endmatrix\right)\, , \, 
   \bar C_{2,\omega} =  
   \left(\matrix
   1&0 \\
   0&0
\endmatrix\right)\, ,\tag2.20
$$
and $\bar D_{2,\omega} = 1 - \epsilon^2\lambda^2$.
Hence, the claim is true for $k = 1, 2$.  
We expand $D_{j,n}$ on the top row, and get the recursion relation
$$
  D_{j,n} =\left(2-\epsilon^2\lambda^2\right)D_{j-1,n} - D_{j-2,n}, 
  \quad 2<j\leq n \tag2.21
$$
Since we are interested in $\omega\in\n$, we will consider the 
cases for which 
$4 -
\left(2-\lambda^2
\epsilon^2\right)^2 > 0$.  
To solve the
difference equation, we substitute $D_{j,n} = Aa^{j-1}$ into (2.21) 
and get
$a^2 - (2-\lambda^2\epsilon^2)a + 1 = 0$ 
with solutions
$$a_{\pm} = \dfrac{\left(2-\lambda^2\epsilon^2\right) \pm i\sqrt{4 -
\left(2-\lambda^2
\epsilon^2\right)^2} }{2}.
\tag2.22$$
Both solutions $a_{\pm}$ have norm 1. Thus,
we can denote $a_{\pm} = e^{i\pm\theta}$, where 
$\theta = arg\left(a_{+}\right) > 0$.  Hence,  
$$
D_{j,n} =
A^{+}exp\left\{i\left(j-1\right)\theta\right\} +
A^{-}exp\left\{-i\left(j-1\right)\theta\right\}. \quad
\tag2.23$$
Solving for the initial conditions
$$\align
  {}&D_1 = 1 = A^{+} + A^- \tag2.24\\
  {}&D_2 = 1 - \epsilon^2\lambda^2 
    = A^+ a_+ + A^- a_-,
\endalign   
$$ 
we get 
$$A^{\pm}=\dfrac{1}{2}\pm
i\left(\dfrac{\lambda\epsilon}
{2\sqrt{4-\lambda^2\epsilon^2}}\right).\tag2.25$$

We now proceed to get an estimate for $\theta$.
By definition of $\theta$, $\cos\theta = 1 - \dfrac{\lambda^2\epsilon^2}{2}$.
After expanding
the left-hand side about $0$, we get
$1-\dfrac{\,\theta^2}{2}\cos\eta=1-\dfrac{\lambda^2\epsilon^2}{2},
\quad 0 < \eta\leq\theta.$  To estimate $\theta$ by $\lambda\epsilon$,
we write $\theta = \lambda\epsilon + \phi$, and obtain
$$\align
{}&1-\dfrac{\,\left(\lambda\epsilon + \phi\right)^2}{2}\cos\eta=
1-\dfrac{\lambda^2\epsilon^2}{2},\quad\Rightarrow \tag2.26 \\
{}&\phi = -\lambda\epsilon\pm
\dfrac{\lambda\epsilon}{\sqrt{\cos\eta}},\quad\Rightarrow\quad
\phi = -\lambda\epsilon+
\dfrac{\lambda\epsilon}{\sqrt{\cos\eta}}\, ,
\endalign
$$
where the last implication is due to $\theta > 0$.
Thus, $\theta = \lambda\epsilon
-\lambda\epsilon\left(1 -
\dfrac{1}{\sqrt{\cos\eta}}\right)$.

By $*$-transforming the above and setting $j = k, n = \omega,
2 < k\leq \omega, \omega\in\n$,
we get
$$\epsilon=\dfrac{t}{\omega}\approx 0,\quad \theta,\eta\approx 0,\quad
1-\dfrac{1}{\sqrt{{}^*\!\!\cos\eta}}\approx 0, \quad
A^{+}\approx{}A^{-}\approx\dfrac{1}2.\tag2.27$$
Equation (2.27) and (2.23) implies that 
$$\align
  {}&\bar D_{k,\omega}\approx
     {}^*\!\!\cos\left\{\left(k-1\right)\theta\right\}
     = \tag2.28\\
  {}&{}^*\!\!\cos\left\{\left(k\lambda\epsilon - \theta - k\lambda\epsilon
      \left(1 -
      \dfrac{1}{\sqrt{{}^*\!\!\cos\eta}}\right)\right)\right\}
      \approx {}^*\!\!\cos\left(k\lambda\epsilon\right)\, .
\endalign
$$
In the last $\approx$ in (2.28),
we used the fact that the cosine function is uniformly
continuous, which translates into ${}*\!\!\cos{}x\approx{}^*\!\!\cos{}y$
whenever $x\approx{}y$ in the language of Nonstandard Analysis. \qed
\enddemo

With the aid of Proposition 2.5, we can show the following:
\proclaim{\bf Theorem 2.6} $\epsilon{}det\,\bar S_{\omega}$ is infinitesimally
close to
$\dfrac{\sin\left(\lambda{}t\right)}{\lambda}.$
\endproclaim

\demo{Proof}  As before, we will use bars to denote the
 $*$-transform of determinate
 of matrices. Let $S_{j,n} = det\left\|A_{j,n} -
    \epsilon^2\lambda^2{}B_{j,n}\right\|$, where 
$A_{j,n}$ is $j$ by $j$ matrix
as defined in $(2.19)$, $B_{j,n}$ is the 
$j$ by $j$ identity matrix,  
$\epsilon = \dfrac {t}{n}, \text{ and }
1\leq j\leq{}n$.  Notice that $S_{n,n}
=det{}\,||S_n||$.  Expanding $S_{j,n}
\text{ and } D_{j,n}$ on the bottom row, we get the recursion relation
$S_{j,n} = D_{j,n} + \left(1-\epsilon^2\lambda^2\right)S_{j-1,n}$, or
equivalently, $S_{j,n} - S_{j-1,n} = D_{j,n} -
\epsilon^2\lambda^2S_{j-1,n}$. 
Summing the last equality gives 
$$\align
  {}&S_{n,n} - S_{1,n} = \sum_{j = 2}^n D_{j,n} -
     \epsilon^2\lambda^2\sum_{j=2}^n S_{j-1,n}, 
     \quad\Rightarrow \tag2.29\\
  {}&\epsilon{}S_{n,n} =
     \epsilon{}S_{1,n} + \sum_{j = 2}^n \epsilon{}D_{j,n} -
     \epsilon^2\lambda^2\sum_{j=2}^n \epsilon{}S_{j-1,n}.
\endalign
$$
From the recursion relation, we also get that for $j\geq 3$, 
$|S_{j-1,n}| \leq\left(\sum_{k=2}^{j-1} |D_{k,n}|\right) + |S_{1,n}|$.

We now $*$-transform the second equation in (2.29) 
and write from proposition (2.5) 
$\bar D_{m,\omega} =
{}^*\!\!\cos\left(m\epsilon\lambda\right) + h_m$,
where $h_m$ is infinitesimal.  We
get
$$
\epsilon{}det\,\bar S_{\omega} = \epsilon\bar S_{1,\omega} +
\sum_{m = 1}^{\omega}
\epsilon{}^*\!\!\cos\left(\frac{mt\lambda}{\omega}\right) -
\epsilon{}^*\!\!\cos\left(\epsilon\lambda\right) +
\epsilon\sum_{m = 2}^{\omega} h_m -
\epsilon^2\lambda^2\sum_{m=2}^{\omega} \epsilon\bar S_{m-1,\omega}.
\tag2.30
$$
The set $\left\{h_m| 1 \leq m \leq \omega\right\}$ is an internal set,
so $\max\limits_{1 \leq m \leq \omega}\!\!h_m \in
\left\{h_m| 1\leq m \leq \omega\right\}$ and it is an infinitesimal.  
Thus, $\epsilon\sum_{m=2}^{\omega}h_m\leq\left(\dfrac{t}{\omega}\right)
\omega\max\limits_{1 \leq m \leq \omega}\!\!h_m{}\approx 0$.  From
the bound on $|S_{j-1,n}|$, we get
$$\align
  {}&|\epsilon^2\lambda^2\sum_{m=2}^{\omega} \epsilon\bar S_{m-1,\omega}|
     \leq\epsilon^3\lambda^2|\bar S_{1,\omega}| + 
     \epsilon^2\lambda^2\left\{\sum_{m=3}^\omega\epsilon
     \left[\left(\sum_{k=2}^{m-1}
     |\bar D_{k,\omega}|\right)
      + |\bar S_{1,\omega}|\right]\right\} < \tag2.31\\
  {}&\epsilon^3\lambda^2|\bar S_{1,\omega}| + 
     \epsilon^2\lambda^2\left[
     \left(\sum_{m=2}^\omega 2t\right)
     + t|\bar S_{1,\omega}|\right] \approx 0.
\endalign
$$
Finally,
using the limit of Riemann sums in the language of Nonstandard Analysis, we
have $\sum_{m = 1}^{\omega}
\epsilon{}^*\cos\left(\frac{mt\lambda}{\omega}\right)
\approx
\int_0^t \cos\left(\lambda{}s\right)\,ds =
\dfrac{\sin\left(\lambda{}t\right)}{\lambda}
$.  Since the other 2 terms in (2.30) 
are also infinitesimals, the result follows.\qed
\enddemo

We are now ready to derive the classical amplitude 
from the exponential in
(2.18) by
using results from theorem 2.6.
\proclaim{\bf Proposition 2.7} The exponential in (2.18)
satisfies the following
$$     \text{exp}\left[\dfrac{im}{\hbar\epsilon}
        \left(\bar w^t\bar T_\omega\bar w -
        \bar \rho^t\bar S_\omega^{-1}\bar \rho\right)\right]\approx
\text{exp}\left\{\dfrac{im}{\hbar}\dfrac{\lambda}{\sin\lambda{}t}
\left[\left(q_0^2 + q^2\right)\cos\lambda{}t - 2qq_0\right]
\right\},\tag2.32$$
where 
$q_0 = x_0, \text{ and } q = x_{\omega + 1}$
\endproclaim

\demo{Proof}
   From the definition of $\rho, w, T_n \text{ and } S_{n}^{-1}$ in
   $(2.12) - (2.14)$, 
  we multiply out $\rho^t{}S_{n}^{-1}{}\rho$ and express it in terms of
  $w^tT_nw$. 
$$\align
  {}&\rho^t{}S_{n}^{-1}{}\rho = 
     \left(\hat w^tS_n - \hat{\hat w}^t\right)S_n^{-1}
     \left(S_n\hat w - \hat{\hat w}\right) = 
     \left(\hat w^t - \hat{\hat w}^tS_n^{-1}\right)
     \left(S_n\hat w - \hat{\hat w}\right) = \tag2.32\\
  {}&\hat w^tS_n\hat w - \hat w^t\hat{\hat w} - 
     \hat{\hat w}^t\hat w + 
     \hat{\hat w}^tS_n^{-1}\hat{\hat w} = 
     w^tT_nw - w_0^2 - \left(1 - \epsilon^2\lambda^2\right)
     w_{n+1}^2 + \\
     {}&w_0^2\left(\bar S_{n}^{-1}\right)_{11} + 
     w_0w_{n + 1}\left(\bar S_{n}^{-1}\right)_{1n} + 
     w_0w_{n + 1}\left(\bar S_{n}^{-1}\right)_{n1} +
     w_{n + 1}^2\left(\bar S_{n}^{-1}\right)_{nn} \\
\endalign
$$
After
  $*$-transforming (2.32), we get
 $$\align {}&\dfrac{1}{\epsilon}
    \left(\bar w^t\bar T_\omega\bar w -
    \bar \rho^t\bar S_\omega^{-1}\bar \rho\right) = 
    \dfrac 1{\epsilon}[
    \bar w_0^2 + \left(1 - \epsilon^2\lambda^2\right)
    \bar w_{\omega + 1}^2 -
    \bar w_0^2\left(\bar S_{\omega}^{-1}\right)_{11} - \tag2.33\\
   {}&\bar w_{\omega + 1}^2\left(\bar S_{\omega}^{-1}\right)_
      {\omega\omega} -
    \bar w_0\bar w_{\omega + 1}\left(\bar S_{\omega}^{-1}\right)_{1\omega} -
    \bar w_0\bar w_{\omega + 1}
    \left(\bar S_{\omega}^{-1}\right)_{\omega{}1}] = \\
  {}&\dfrac {1}{\epsilon}\left[q_0^2\left(1-\dfrac{\bar S_{\omega - 1, \omega}}
    {\bar S_{\omega,\omega{}}}\right) +
    q^2\left(1-\dfrac{\bar S_{\omega - 1, \omega}}
    {\bar S_{\omega,\omega{}}}\right) -
    2qq_0\dfrac{\left(-1\right)^{\omega + 1}\left(-1\right)^{\omega - 1}}
             {\bar S_{\omega,\omega}} - \epsilon^2\lambda^2q^2\right] = \\
  {}&\dfrac {1}{\epsilon{}det\bar S_{\omega}}
        \left[q_0^2\left(\bar S_{\omega ,\omega} -
        \bar S_{\omega - 1, \omega}\right) +
        q^2\left(\bar S_{\omega ,\omega} -
        \bar S_{\omega - 1, \omega}\right) - 2qq_0
        \right] - \epsilon\lambda^2q^2\approx \\
  {}& \dfrac{\lambda}{\sin\lambda{}t}
        \left[\left(q_0^2+q^2\right)\cos\lambda{}t - 2qq_0\right]\, .
\endalign $$
In the third line above, we used $*$-Cramer's rule 
and our previous definition
of $\bar S_{k,\omega}$. The factor 
$\left(-1\right)^{\omega + 1}$
comes from a cofactor expansion in the numerator of
Cramer's rule for $\left(S_{\omega}^{-1}\right)_{1\omega}$ and
$\left(S_{\omega}^{-1}\right)_{\omega{}1}$; the 
factor $\left(-1\right)^{\omega - 1}$ comes
from the determinate of a triangular matrix with 
$-1$'s along the diagonal after the latter expansion.  
In the fourth line, we used results from
Theorem 2; namely, $\bar S_{\omega,\omega} =
det\,\bar S_{\omega},
\,\epsilon{}det\,\bar S_{\omega}\approx\dfrac{\sin\lambda{}t}{\lambda}
, \text{ and } \bar S_{\omega ,\omega} - \bar S_{\omega - 1, \omega}\approx{}
\bar D_{\omega , \omega}\approx\cos\lambda{}t$. \qed
\enddemo

Notice that in (2.33), the end result 
dependent only on the endpoints of the path $w$.  
It does not matter which path is choosen as long as
it starts at $q$ and ends at $q_0$.  Hence, it is not
necessary to use the classical path $x^{cl}$ to do
the computation.  
\proclaim{Theorem 2.8} For $t < \dfrac{\pi}{\lambda}$,
the one dimensional harmonic oscillator standard functional integral
is given by
$$\left(\dfrac{m}{2\pi{}i\hbar}\right)^{\frac{1}2}
\sqrt{\dfrac{\lambda}
{\sin\lambda{}t}}
exp\left\{\dfrac{im}{\hbar}\dfrac{\lambda}{\sin\lambda{}t}
\left[\left(q_0^2 + q^2\right)\cos\lambda{}t - 2qq_0\right]
\right\}.\tag2.34$$
\endproclaim

\demo{Proof} This follows from 
corollary 2.4, theorem 2.6, and proposition 2.7. \qed
\enddemo

\proclaim{\bf Corollary 2.9} 
For the $d$-dimensional harmonic oscillator standard
functional integral, we have
$$\align
      {}&st\Bigg\{\int\limits_{{}^*\Bbb R^{d\omega}}
        \fac{d}{\omega}*\tag2.35\\
      {}&\text{exp}\left[\summ{\omega}{x_j}{x_{j-1}}
       {-\dfrac{m}2\lambda^2 x_j^2}\right]
      dx_1\dots dx_{\omega}\Bigg\} = \\
      {}&\left(\dfrac{m}{2\pi{}i\hbar}\right)^{\frac{d}2}
      \left(\dfrac{\lambda}
      {\sin\lambda{}t}\right)^{\frac{d}{2}}
      \text{exp}\left\{\dfrac{im}{\hbar}\dfrac{\lambda}{\sin\lambda{}t}
      \left[\left({\vec {q_0}}^2 + {\vec q}^2\right)
       \cos\lambda{}t -
       2{\vec q}{\vec {q_0}}\right]
      \right\}.
  \endalign $$
\endproclaim

\demo{Proof} Follows from factoring (2.35) into products
of one dimensional harmonic oscillator standard
functional integrals and theorem 2.8. \qed
\enddemo

\enddocument